\UseRawInputEncoding
\pdfoutput=1
\documentclass[a4paper,11pt]{article}
\pdfoutput=1 
\usepackage{jheppub} 
\usepackage[T1]{fontenc} 
\RequirePackage[displaymath]{lineno} 

\usepackage{epsfig}
\usepackage{graphicx}
\usepackage{dcolumn}
\usepackage{bm}
\usepackage{ltablex,booktabs}
\usepackage{overpic}
\usepackage{subfigure}
\usepackage{float}
\usepackage{color}
\usepackage{amsmath}
\usepackage{mathcomp}
\usepackage{mathrsfs}
\usepackage{multirow}
\usepackage{rotating}
\usepackage{amssymb}
\usepackage{gensymb}
\usepackage{tabularx}
\usepackage{epstopdf}

\title{Measurement of absolute branching fractions of $D_s^+$ hadronic decays}
\collaborationImg{\includegraphics[width=0.15\textwidth, angle=90]{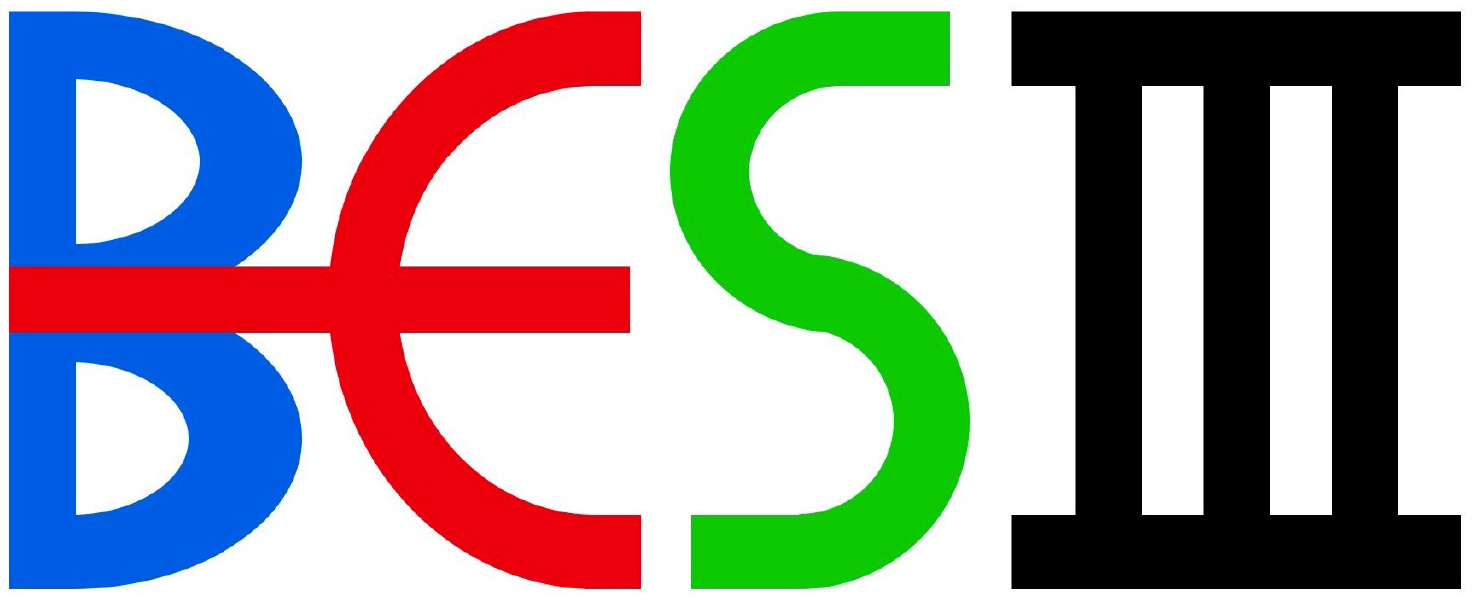}}
\collaboration{The BESIII Collaboration}

\date{\today}
\abstract{Using $e^+ e^-$ collision data collected at the BESIII detector at center-of-mass energies between 4.128 and 4.226 GeV, corresponding to an integrated luminosity of $7.33~{\rm fb}^{-1}$,
we determine the absolute branching fractions of fifteen hadronic $D_s^{+}$ decays with a double-tag technique.
In particular, we make precise measurements of the branching fractions $\mathcal{B}(D_s^+ \to K^+ K^- \pi^+)=(5.49 \pm 0.04 \pm 0.07)\%$, $\mathcal{B}(D_s^+ \to K_S^0 K^+)=(1.50 \pm 0.01 \pm 0.01)\%$ and $\mathcal{B}(D_s^+ \to K^+ K^- \pi^+ \pi^0)=(5.50 \pm 0.05 \pm 0.11)\%$, 
where the first uncertainties are statistical and the second ones are systematic. 
The \emph{CP} asymmetries in these decays are also measured and all are found to be compatible with zero.}

\arxivnumber{}
\begin{document}
\maketitle
\flushbottom

\section{Introduction}

Hadronic $D_s^+$ decays play an important role in both charm and beauty physics.
Precise measurements of the absolute branching fractions~(BF) of hadronic $D_s^+$ decays provide useful information to understand the underlying decay mechanism and help improve theoretical
models~\cite{
PhysRevD.105.033006,
PhysRevD.105.093006,
PhysRevD.100.093002,
Qin:2013tje,
Li:2012cfa,
PhysRevD.84.074019,
PhysRevD.81.014026,
Bhattacharya:2008ke}.
The $D_s^+ \to K^+ K^- \pi^+$, $D_s^+ \to K_S^0 \pi^+$, and $D_s^+ \to K^+ K^- \pi^+ \pi^0$ decays are not only used as the reference modes for measurements of relative BFs of $D_s^+$ decays, but also used to measure the modulus of the Cabibbo-Kobayashi-Maskawa matrix element $|V_{cb}|$ in decays $B_s^0 \to D_s^- X$~\cite{PhysRevD.101.072004}.
Furthermore, searching for \emph{CP} asymmetries in hadronic $D_s^{\pm}$ decays allows a more comprehensive understanding of \emph{CP} violation in the charm sector.
Therefore, precision measurements of the absolute BFs of hadronic $D_s^+$ are an important component of the experimental and theoretical heavy-flavor physics program.

In this paper,
we perform the measurements of the absolute BFs of fifteen $D_s^+$ decays: $K^{0}_{S} K^{+}$, $K^{+} K^{-} \pi^{+}$, $K^{0}_{S} K^{+} \pi^{0}$, $K^{0}_{S} K^{0}_{S} \pi^{+}$, $K^{+} K^{-} \pi^{+} \pi^{0}$, $K^{0}_{S} K^{+} \pi^{+} \pi^{-}$, $K^{0}_{S} K^{-} \pi^{+} \pi^{+}$, $\pi^{+} \pi^{+} \pi^{-}$, $\pi^{+} \eta$, $\pi^{+} \pi^{0} \eta$, $\pi^{+} \pi^{+} \pi^{-} \eta$, $\pi^{+} \eta'$, $\pi^{+} \pi^{0} \eta'$, $K^{0}_{S} \pi^{+} \pi^{0}$, $K^{+} \pi^{+} \pi^{-}$,   
based on a sample of $e^+e^-$ annihilation data corresponding to an integrated luminosity of $7.33~{\rm fb^{-1}}$ taken at the center-of-mass  energies in the interval $\sqrt{s}=4.128-4.226~{\rm GeV}$ with the BESIII detector~\cite{Ablikim:2009aa}. 
Thirteen of these decay modes have been previously measured by the CLEO-c collaboration~\cite{cleo_2013}. 
Moreover, \emph{CP} asymmetries for these decays are measured in this paper. 
Throughout this paper, charge-conjugated processes are implied except in \emph{CP} asymmetry measurements.

\section{Measurement technique}

The double-tag (DT) method~\cite{PhysRevLett.56.2140} is employed to obtain clean signal samples of $e^+ e^- \to D_s^{*\pm} D_s^{\mp} \to (\gamma,~ \pi^0) D_s^+ D_s^-$ in the following analyses. 
The transition photon or $\pi^0$ is not reconstructed.
In this analysis, 
a single-tag~(ST) candidate requires only one of the $D_s^{\pm}$ mesons to be reconstructed via a hadronic decay, 
and a DT candidate has both $D_s^+$ and $D_s^-$ mesons reconstructed via hadronic decays. 
Considering two ST modes,
$D_s^+ \to i$ and $D_s^- \to \bar{j}$,
and one DT mode $D_s^+ \to i$, $D_s^- \to \bar{j}$, 
the BF for the $i$ and $\bar{j}$ decays are $\mathcal{B}_i$ and $\mathcal{B}_{\bar{j}}$.
We assume \emph{CP} violation is negligible while determining the BF~($\mathcal{B}_j=\mathcal{B}_{\bar{j}}$),
then we have:
\begin{eqnarray} 
\begin{aligned}
& Y_i= N^{D_s^+ D_s^-} \mathcal{B}_i \epsilon_i  \,,\\
& Y_{\bar{j}}= N^{D_s^+ D_s^-} \mathcal{B}_{\bar{j}}   \epsilon_{\bar{j}}\,,\\
& Y_{i \bar{j}} = N^{D_s^+ D_s^-} \mathcal{B}_i \mathcal{B}_{\bar{j}} \epsilon_{i \bar{j}}   \,,\\
\end{aligned}
\label{eq:bf}
\end{eqnarray}
where 
$\epsilon_i$ and $\epsilon_{\bar{j}}$ are the ST efficiencies, 
$\epsilon_{i\bar{j}}$ is the DT efficiency, 
$Y_i$ and $Y_{\bar{j}}$ are the expected ST yields, 
$Y_{i \bar{j}}$ is the expected DT yield, 
and $N^{D_s^+ D_s^-}$ is the number of $D_s^+ D_s^-$ pairs. 

For ST and DT yields, we calculate the expected yields based on Eq.~(\ref{eq:bf}) using $B_i$, $B_{\bar{j}}$ and $N^{D_s^+ D_s^-}$, and perform a maximum likelihood fit to the obtained ST and DT yields. Through the fit, we obtain the results of the parameters, considering both statistical and systematic uncertainties.

We constrain $\mathcal{B}_i$ to be the same for different final states involving $\eta$~($\eta'$) in the intermediate state, 
taking into account different detection efficiencies and BFs of the $\eta$~($\eta'$) meson. 

We analyze 
fifteen decay modes~(nineteen final states) and seven data sample groups~(see table~\ref{energe}), 
leading to a total of 266~($19 \times 2 \times 7$) ST yields,
2527~($19 \times 19 \times 7$) DT yields, 
fifteen BFs 
and seven $N^{D_s^+ D_s^-}$ values.

We derive the CP asymmetry for each decay mode by:
\begin{eqnarray}
\begin{aligned}
\mathcal{A}_{\emph{CP},i}=\frac
{  N_i/\epsilon_i-N_{\bar{i}}/\epsilon_{\bar{i}}}
{  N_i/\epsilon_{i}+N_{\bar{i}}/\epsilon_{\bar{i}}},
\end{aligned}
\end{eqnarray}
where $N_{i}$~($N_{\bar{i}}$) is the obtained yield for ST mode $i$~($\bar{i}$).

\section{Detector and data sets}

The BESIII detector~\cite{Ablikim:2009aa} records symmetric $e^+e^-$ collisions 
provided by the BEPCII storage ring~\cite{Yu:IPAC2016-TUYA01}
in the center-of-mass energy range from 2.0 to 4.95~GeV,
with a peak luminosity of $1 \times 10^{33}\;\text{cm}^{-2}\text{s}^{-1}$ 
achieved at $\sqrt{s}=3.77\;\text{GeV}$. 
The cylindrical core of the BESIII detector covers 93\% of the full solid angle and consists of a helium-based
 multilayer drift chamber~(MDC), a plastic scintillator time-of-flight
system~(TOF), and a CsI(Tl) electromagnetic calorimeter~(EMC),
which are all enclosed in a superconducting solenoidal magnet
providing a 1.0~T magnetic field~\cite{detvis}.
The solenoid is supported by an
octagonal flux-return yoke with resistive plate counter muon
identification modules interleaved with steel. 
The charged-particle momentum resolution at $1~{\rm GeV}/c$ is
$0.5\%$, and the 
${\rm d}E/{\rm d}x$
resolution is $6\%$ for electrons
from Bhabha scattering. The EMC measures photon energies with a
resolution of $2.5\%$~($5\%$) at $1$~GeV in the barrel~(end cap)
region. The time resolution in the TOF barrel region is 68~ps, while
that in the end cap region is 110~ps. The end cap TOF
system was upgraded in 2015 using multigap resistive plate chamber
technology, 
providing a time resolution of 60~ps~\cite{etof1,etof2,etof3}. 
About 84\% of the data used in this paper benefits from this upgrade.  
 
Data samples corresponding to a total integrated luminosity of 7.33 f$\rm b^{-1}$ are used in this paper. 
The integrated luminosities for the individual center-of-mass energies~\cite{Li:2021iwf} are given in table~\ref{energe}.
Data samples of $\sqrt{s}=4.128$~GeV and $\sqrt{s}=4.157$~GeV are merged into one group due to their low statistics.  
Since the cross section of $e^+ e^- \to D_s^{* \pm} D_s^{\mp}$ production in $e^+ e^-$ annihilation is about twenty times larger than the one of $e^+ e^- \to D_s^+ D_s^-$~\cite{BESIII:2024zdh} in this energy region, 
the signal events discussed in this paper are selected from the process $e^+ e^- \to D_s^{* \pm} D_s^{\mp}$. 
 
\begin{table}[!htbp]
\renewcommand\arraystretch{1.25}
\centering

 \begin{tabular}{c c c}
 \hline
 $\sqrt{s}$~(GeV) & $\mathcal{L}_{\rm int}$~(pb$^{-1}$) & $M_{\rm rec}$~(GeV/$c^2$)\\
 \hline
  4.128 &  401.5                      & [2.060, 2.150] \\
  4.157 &  408.7                      & [2.054, 2.170] \\
  4.178 & $3189.0\pm0.2\pm31.9$ & [2.050, 2.180] \\
  4.189 &  $570.0\pm0.1\pm2.2$  & [2 048, 2.190] \\
  4.199 &  $526.0\pm0.1\pm2.1$  & [2.046, 2.200] \\
  4.209 &  $572.1\pm0.1\pm1.8$  & [2.044, 2.210] \\
  4.219 &  $569.2\pm0.1\pm1.8$  & [2.042, 2.220] \\
  4.226 & $1100.9\pm0.1\pm7.0$  & [2.040, 2.220] \\
  \hline
 \end{tabular}
  \caption{The integrated luminosities~($\mathcal{L}_{\rm int}$) and the requirements on $M_{\rm rec}$ for various center-of-mass energies~\cite{s1,l}. 
   The first and second uncertainties are statistical and systematic, respectively. 
   The definition of $M_{\rm rec}$ is given in Eq.~(\ref{eq:mrec}). 
   The integrated luminosities for data samples of $\sqrt{s} = 4.128~\rm{GeV}$ and $\sqrt{s} = 4.157~\rm{GeV}$ are estimated by using online monitoring information.}
    \label{energe}
\end{table}

To determine the detection efficiencies and estimate backgrounds,
we produce and analyze {\sc geant4}-based~\cite{geant4} Monte Carlo~(MC) simulation samples for all data sets listed in
table~\ref{energe}, with sizes that are 40 times the integrated luminosity of data. The MC samples are produced using known
decay rates~\cite{PDG} and correct angular distributions by two event generators, {\sc evtgen}~\cite{ref:evtgen} for
charm ($D_s^{* \pm}$, $D_s^{\pm}$, $D^{*0(\pm)}$, and $D^{0(\pm)}$) and charmonium decays and {\sc kkmc}~\cite{kkmc} for continuum processes.
The samples consist of $e^+ e^- \to D \bar{D}$, $D^*D$, $D^* D^*$, $D_s D_s$, $D_s^* D_s$, $D_s^* D_s^*$, $DD^* \pi$, $DD\pi$, $q\bar{q}\,(q=u,d,s)$, $\gamma {\mathit J / \psi}$, $\gamma\,\psi(3686)$, and $\tau^+ \tau^-$. 
For the fifteen $D_s^+$ signal processes, twelve multi-body processes are
generated based on amplitude models~\cite{mode_401,mode_402,mode_403,mode_404,mode_406,mode_421,mode_441,mode_442,mode_461,mode_501,mode_502}\footnote{The amplitude model for the $K_S^0 K^+ \pi^+ \pi^-$ mode is taken from an unpublished BESIII internal results, which are on going.}, while the three two-body processes are modeled with a uniform phase-space distribution.
Charmonium decays that are not accounted for by exclusive measurements are simulated by {\sc lundcharm}~\cite{ref:lundcharm}.
All MC simulations include the effects of initial-state radiation (ISR) and final-state radiation (FSR).
We simulate ISR with ConExc~\cite{conexc} for $e^+ e^- \to c \bar{c}$ events within the framework of {\sc evtgen},
and with {\sc kkmc} for non-charm continuum processes.
The simulation models the beam energy spread in the $e^+e^-$ annihilations with the generator {\sc kkmc}.
FSR from charged final state particles is incorporated using {\sc photos}~\cite{photos}.


\section{Event selection}
\label{Event-selection}

The $D_s^{\pm}$ candidates are constructed from combinations of $\pi^{\pm}$, $K^{\pm}$, $K_S^0$, $\eta$, $\eta'$, $\rho(770)^0$, and $\gamma$ candidates in nineteen final states.
The $D_s^+ \to \pi^+ \eta$ decay is reconstructed via two distinct final states, $D_s^+ \to \pi^+ \eta_{\gamma \gamma}$ and $D_s^+ \to \pi^+ \eta_{\pi^+ \pi^- \pi^0}$.  
The $D_s^+ \to \pi^+ \eta'$ is reconstructed in three final states $D_s^+ \to \pi^+ \eta'_{\pi^+ \pi^- \eta_{\gamma \gamma}}$, $D_s^+ \to \pi^+ \eta'_{\pi^+ \pi^- \eta_{\pi^+ \pi^- \pi^0}}$, and $D_s^+ \to \pi^+  \eta'_{\gamma \rho}$.
The $D_s^+ \to \pi^+ \pi^0 \eta'$ is reconstructed in two final states which are $D_s^+ \to \pi^+ \pi^0 \eta'_{\pi^+ \pi^- \eta_{\gamma \gamma}}$, and $D_s^+ \to \pi^+ \pi^0 \eta'_{\gamma \rho}$.
Here, the subscripts on $\eta_{\gamma \gamma}$, $\eta_{\pi^+ \pi^- \pi^0}$, $\eta'_{\pi^+ \pi^- \eta}$, and $\eta'_{\gamma \rho}$ indicate the reconstructed decay modes $\eta \to \gamma \gamma$, $\eta \to \pi^+ \pi^- \pi^0$, $\eta' \to \pi^+ \pi^- \eta$ and $\eta' \to \gamma \rho$,
where $\rho$ denotes $\rho(770)^0$.

All charged tracks detected in the MDC are required to be within a polar angle~($\theta$) range of $|\rm{cos}\theta|<0.93$, where $\theta$ is defined with respect to the $z$-axis,
which is the symmetry axis of the MDC. 
For charged tracks not originating from $K_S^0$ decays, the distance of closest approach to the interaction point~(IP) 
must be less than 10\,cm
along the $z$-axis, $|V_{z}|$,  
and less than 1\,cm
in the transverse plane, $|V_{xy}|$.
Particle identification~(PID) for charged tracks combines measurements of the d$E$/d$x$ in the MDC and the flight time in the TOF to form likelihoods $\mathcal{L}~(h)~(h=K,\pi)$ for each hadron $h$ hypothesis.
The charged kaons and pions are identified by requiring $\mathcal{L}~(K)>\mathcal{L}~(\pi)$ and $\mathcal{L}~(\pi)>\mathcal{L}~(K)$, respectively.

The $K_{S}^0$ candidates are reconstructed from pairs of oppositely charged tracks satisfying $|V_{z}|<$ 20~cm.
The two charged tracks are assigned
as $\pi^+\pi^-$ without imposing the above PID criteria. 
The quality of the vertex fits is ensured by a requirement of $\chi^2<100$. 
The invariant mass of the $\pi^+ \pi^-$ pair is required to be within $[0.487,~0.511]~{\rm GeV}/c^2$.
For the $D_s^+ \to K_S^0 K_S^0 \pi^+$ and $D_s^+ \to K_S^0 \pi^+ \pi^0$ modes, 
to avoid the peaking background from $D_s^+ \to \pi^+ \pi^+ \pi^+ \pi^- \pi^-$ and $D_s^+ \to \pi^+ \pi^+ \pi^- \pi^0$ modes,  
the decay length from the IP of the $K^0_S$ candidate is required
to be greater than twice the resolution. 

Photon candidates are identified using showers in the EMC.  The deposited energy of each shower must be more than 25~MeV in the barrel region~($|\rm{\cos}\theta|< 0.80$) and more than 50~MeV in the end cap region~($0.86 <|\rm{\cos}\theta|< 0.92$).
To exclude showers that originate from
charged tracks,
the angle subtended by the EMC shower and the position of the closest charged track at the EMC
must be greater than 10 degrees as measured from the IP. 
To suppress electronic noise and showers unrelated to the event, the difference between the EMC time and the event start time is required to be within [0, 700]\,ns.

The $\pi^0$ and $\eta_{\gamma \gamma}$ candidates are reconstructed from photon pairs with invariant masses in the ranges $[0.115,~0.150]$~GeV/$c^{2}$ and $[0.490,~0.580]$~GeV/$c^{2}$, respectively.
We require that at least one photon comes from the barrel region of the EMC to improve their invariant mass resolutions. 
A kinematic fit constraining the invariant mass of the selected photon pair to the known $\pi^{0}$ or $\eta$ mass~\cite{PDG} is performed, 
and the $\chi^2$ of the kinematic fit is required to be less than 30. 
The $\eta_{\pi^+ \pi^- \pi^0}$ candidates are formed from $\pi^+ \pi^- \pi^0$ combinations with an invariant mass in the range  $[0.530,~0.560]$~GeV/$c^2$. 

For the $\eta^{\prime}$ candidates formed from $\pi^{+}\pi^{-}\eta$ and $\gamma \rho(770)^0$ combinations, 
we require invariant masses to be within the ranges $[0.943,~0.973]$~GeV/$c^{2}$ and $[0.946,~0.970]$~GeV/$c^{2}$, respectively.
Furthermore, 
the $\rho(770)^0$ candidates are formed from $\pi^+ \pi^-$ combinations with an invariant mass within the range $[0.570,~0.970]$~GeV/$c^2$.

For the $D_s^+ \to K^+ \pi^+ \pi^-$ mode, 
we require the $\pi^+ \pi^-$ invariant mass to be outside
of the range $[0.487,~0.511]$~GeV/$c^2$ 
to exclude the $K_S^0 \to \pi^+ \pi^-$ contamination
from the process $D_s^+ \to K_S^0 K^+$. 
Similarly, for the $D_s^+ \to \pi^+ \pi^+ \pi^- \eta$ mode, 
we require the $\pi^+ \pi^- \eta$ invariant mass to be outside
of the range $[0.943,~0.973]$~GeV/$c^2$
to exclude the $\eta^{\prime} \to \pi^+ \pi^- \eta$ contribution
from the process $D_s^+ \to \pi^+ \eta^{\prime}$.   

The invariant masses of $D_s^{\pm}$ candidates~($M_{D_s^{\pm}}$) are required to be in the range of $[1.88,~2.06]~{\rm GeV}/c^2$. 
The recoil mass $M_{\rm rec}$ is defined as

\begin{eqnarray}
\begin{aligned}
	\begin{array}{l r}
	M_{\rm rec}^2 c^4  = \left (\sqrt{s} - \sqrt{|\vec{p}_{D_{s}^{+}}|^{2} c^2+m_{D_{s}^{+}}^{2} c^4}\right)^{2} - \left|\vec{p}_{D_{s}^{+}} \right| ^{2} c^2 \; ,
		\end{array}
\end{aligned}
 \label{eq:mrec}
\end{eqnarray}
where $\vec{p}_{D_{s}^{+}}$ is the three-momentum of the $D_{s}^{+}$ candidate in the $e^+e^-$ center-of-mass frame 
and $m_{D_{s}^{+}}$ is the known $D_{s}^{+}$ mass~\cite{PDG}. 
We use two different requirements on $M_{\rm rec}$: a tighter 
selection, within the interval of $[2.10,~2.13]~{\rm GeV}/c^2$, for $D_s^+ \to \pi^+ \pi^+ \pi^- \eta$, 
$D_s^+ \to \pi^+ \pi^0 \eta_{\gamma \rho}^{\prime}$, and 
$D_s^+ \to K_S^0 \pi^+ \pi^0$ final states for all energy points to reduce background, 
and the looser intervals given in table~\ref{energe} for the remaining final states. 
If there are multiple ST candidates, 
the candidate with the $M_{\rm rec}$ closest to the known $D_s^{*+}$ mass~\cite{PDG} is kept.

The DT $D_s^{\pm}$ candidates are required to pass the same selections as those for ST candidates,  
and we require one of the two $D_s^{\pm}$ candidates to satisfy $M_{\textrm{rec}}>2.10~{\rm GeV}/c^2$. 
If there are multiple combinations in an event, the combination with the average $D_s^{\pm}$ candidate invariant mass~($\overline{m} = (M_{D_s^+}+M_{D_s^-})/2$) closest to the known $D_s^{\pm}$ mass~\cite{PDG} is retained. About 19.7\% (20.1\%) of events in data (simulated samples) contain multiple DT candidates,
and the distributions of the multiplicity of DT candidate show good consistency between data and simulated samples. 
This requirement retains more than 97.7\% of DT signal candidates.


\section{Single-tag and double-tag yields}
\label{section:yields}

\begin{figure}[!htbp]
\centering
\begin{overpic}[width=1.0\textwidth,height=1.25\textwidth]{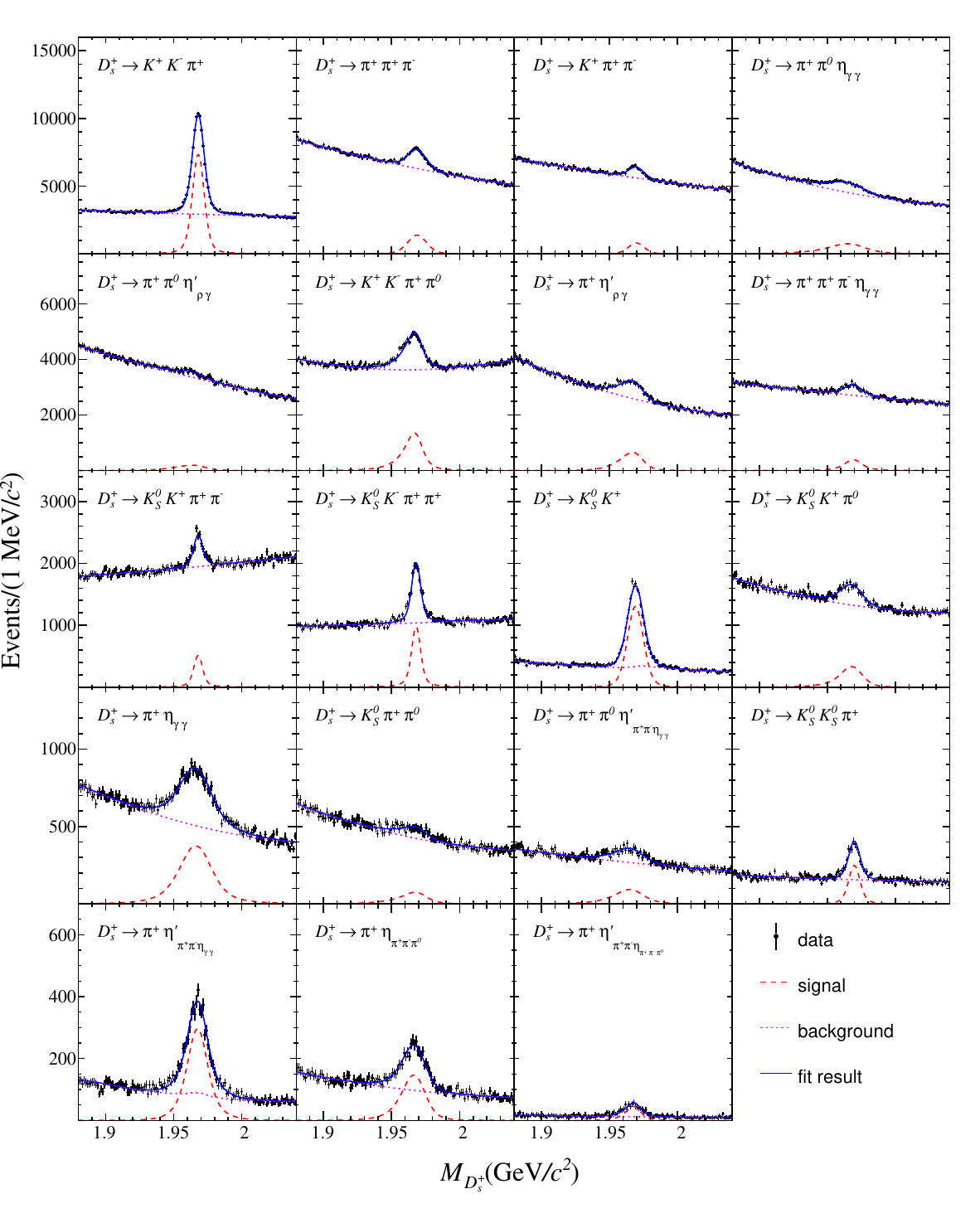}
\end{overpic}
\caption{Fits to the $M_{D_s^+}$ distributions of the ST candidates from the data sample at $\sqrt{s}=4.178~\rm{GeV}$. The points with error bars are data, the blue solid curves are the fit results, the pink dotted curves are the fitted background shapes, and the red dashed curves are the signals. }
\label{fig:st_data_1}
\end{figure}

\begin{figure}[!htbp]
\centering
\begin{overpic}[width=1.0\textwidth,height=1.25\textwidth]{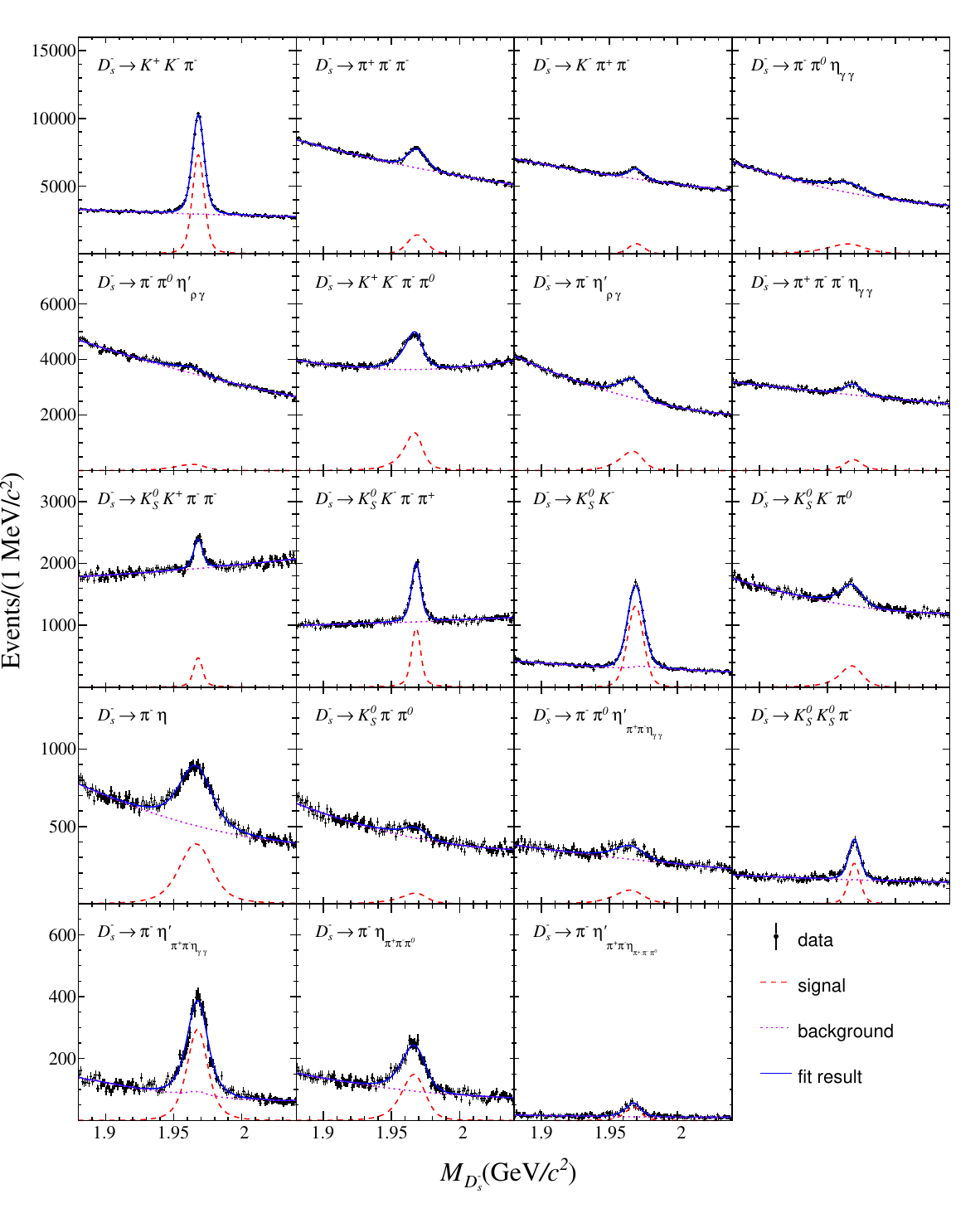}
\end{overpic}
\caption{Fits to the $M_{D_s^-}$ distributions of the ST candidates from the data sample at $\sqrt{s}=4.178~\rm{GeV}$. The points with error bars are data, the blue solid curves are the fit results, the pink dotted curves are the fitted background shapes, and the red dashed curves are the signals. }
\label{fig:st_data_2}
\end{figure}

The ST yields are determined from fits to the $M_{D_s^{\pm}}$ distributions. 
As an example, the fits to the $M_{D_s^+}$ and $M_{D_s^-}$  distributions of the selected ST candidates from the data sample at $\sqrt{s}=4.178~\rm{GeV}$ are shown in figure~\ref{fig:st_data_1} and figure~\ref{fig:st_data_2}, respectively. 
We use the MC-simulated shape convolved with a Gaussian function to describe the signal.
The background is described by a second-order polynomial. 
The MC-simulated shapes of the final states $D^+ \to K_S^0 \pi^+$ and $D_s^+ \to \pi^+ \pi^+ \pi^- \eta_{\gamma \gamma}$  are included as peaking background components in the fits for the final states $D_s^+ \to K_S^0 K^+$ and $D_s^+ \to \pi^+ \eta_{\pi^+ \pi^- \eta_{\gamma \gamma }}'$, respectively. The various ST yields in data at $\sqrt{s}=4.178~\rm{GeV}$ are summarized in table~\ref{tab:Y_4180}.

\begin{table}[!htbp]
\centering
\begin{tabular}{l r@{$\pm$}l l r@{$\pm$}l}
\hline 
Final state  &  \multicolumn{2}{c}{Yield} & Final state  &  \multicolumn{2}{c}{Yield} \\
\hline
$D_{s}^{+} \to K^{0}_{S} K^{+}$
 & 16668~ & ~166 & $D_{s}^{-} \to K^{0}_{S} K^{-}$ & 16739~ & ~166\\
$D_{s}^{+} \to K^{+} K^{-} \pi^{+}$
 & 73252~ & ~379 & $D_{s}^{-} \to K^{+} K^{-} \pi^{-}$ & 73048~ & ~380\\
$D_{s}^{+} \to K^{0}_{S} K^{+} \pi^{0}$
 & 6375~ & ~249 & $D_{s}^{-} \to K^{0}_{S} K^{-} \pi^{0}$ & 6616~ & ~249\\
$D_{s}^{+} \to K^{0}_{S} K^{0}_{S} \pi^{+}$
 & 2546~ & ~79 & $D_{s}^{-} \to K^{0}_{S} K^{0}_{S} \pi^{-}$ & 2686~ & ~80\\
$D_{s}^{+} \to K^{+} K^{-} \pi^{+} \pi^{0}$
 & 24033~ & ~425 & $D_{s}^{-} \to K^{+} K^{-} \pi^{-} \pi^{0}$ & 24295~ & ~426\\
$D_{s}^{+} \to K^{0}_{S} K^{+} \pi^{+} \pi^{-}$
 & 4944~ & ~217 & $D_{s}^{-} \to K^{0}_{S} K^{-} \pi^{+} \pi^{-}$ & 4580~ & ~215\\
$D_{s}^{+} \to K^{0}_{S} K^{-} \pi^{+} \pi^{+}$
 & 9156~ & ~182 & $D_{s}^{-} \to K^{0}_{S} K^{+} \pi^{-} \pi^{-}$ & 8904~ & ~181\\
$D_{s}^{+} \to \pi^{+} \pi^{+} \pi^{-}$
 & 20655~ & ~444 & $D_{s}^{-} \to \pi^{+} \pi^{-} \pi^{-}$ & 20875~ & ~446\\
$D_{s}^{+} \to \pi^{+} \eta_{\gamma \gamma}$
 & 10755~ & ~225 & $D_{s}^{-} \to \pi^{-} \eta_{\gamma \gamma}$ & 11131~ & ~226\\
$D_{s}^{+} \to \pi^{+} \pi^{0} \eta_{\gamma \gamma}$
 & 24551~ & ~649 & $D_{s}^{-} \to \pi^{-} \pi^{0} \eta_{\gamma \gamma}$ & 24050~ & ~648\\
$D_{s}^{+} \to \pi^{+} \pi^{+} \pi^{-} \eta_{\gamma \gamma}$
 & 6140~ & ~319 & $D_{s}^{-} \to \pi^{+} \pi^{-} \pi^{-} \eta_{\gamma \gamma}$ & 6215~ & ~320\\
$D_{s}^{+} \to \pi^{+} \eta_{\pi^+ \pi^- \pi^0}$
 & 2898~ & ~88 & $D_{s}^{-} \to \pi^{-} \eta_{\pi^+ \pi^- \pi^0}$ & 2914~ & ~87\\
$D_{s}^{+} \to \pi^{+} \eta_{\pi^+ \pi^- \eta_{\gamma \gamma}}' $
 & 5287~ & ~99 & $D_{s}^{-} \to \pi^{-} \eta_{\pi^+ \pi^- \eta_{\gamma \gamma}}' $ & 5239~ & ~99\\
$D_{s}^{+} \to \pi^{+} \pi^{0} \eta_{\pi^+ \pi^- \eta_{\gamma \gamma}}'$
 & 2326~ & ~136 & $D_{s}^{-} \to \pi^{-} \pi^{0} \eta_{\pi^+ \pi^- \eta_{\gamma \gamma}}'$ & 2197~ & ~139\\
$D_{s}^{+} \to \pi^{+} \eta'_{\pi^{+}\pi^{-}\eta_{\pi^+ \pi^- \pi^0}}$
 & 624~ & ~33 & $D_{s}^{-} \to \pi^{-} \eta'_{\pi^{+}\pi^{-}\eta_{\pi^+ \pi^- \pi^0}}$ & 609~ & ~33\\
$D_{s}^{+} \to \pi^{+} \eta'_{\rho \gamma}$
 & 13192~ & ~359 & $D_{s}^{-} \to \pi^{-} \eta'_{\rho \gamma}$ & 14042~ & ~362\\
$D_{s}^{+} \to \pi^{+} \pi^{0} \eta'_{\rho \gamma}$
 & 5156~ & ~468 & $D_{s}^{-} \to \pi^{-} \pi^{0} \eta'_{\rho \gamma}$ & 5859~ & ~471\\
$D_{s}^{+} \to K^{0}_{S} \pi^{+} \pi^{0}$
 & 1910~ & ~172 & $D_{s}^{-} \to K^{0}_{S} \pi^{-} \pi^{0}$ & 1703~ & ~171\\
$D_{s}^{+} \to K^{+} \pi^{+} \pi^{-}$
 & 10589~ & ~394 & $D_{s}^{-} \to K^{-} \pi^{+} \pi^{-}$ & 9991~ & ~390\\
\hline
\end{tabular}
\caption{ST yields in data with statistical uncertainties at $\sqrt{s}$=4.178~GeV.}
\label{tab:Y_4180}
\end{table}

Since each ST final state may receive crossfeed background from other ST signal final states,  
we create an efficiency-correction matrix, $\textbf{C}^{\rm ST}$, to describe simultaneously the detection efficiencies (diagonal elements) and the crossfeed probabilities (off-diagonal elements)~\cite{sun2006simultaneous}.
The elements $\textbf{C}^{\rm ST}_{ij}$ are defined to be the probabilities that an event of signal mode $j$ is reconstructed and counted in the yield for mode $i$, as determined using MC samples. 

\begin{figure}[!htbp]
\centering
\begin{overpic}[width=0.6\textwidth,height=0.6\textwidth]{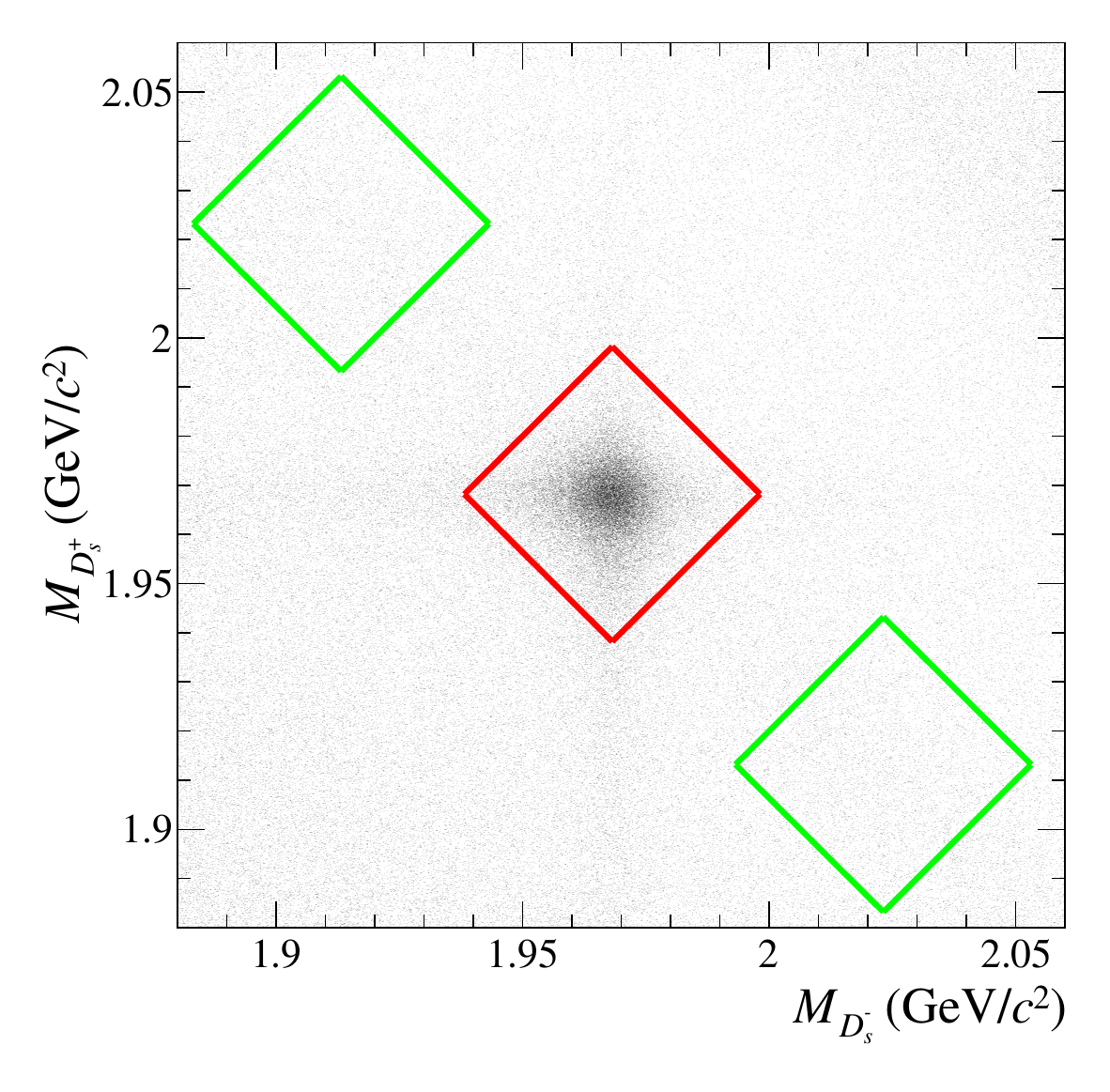}
\end{overpic}
\caption{Invariant mass of the $D_s^+$ candidate versus invariant mass of the $D_s^-$ candidate for all 361 DT final states and seven data samples. The squares show the signal region~(red) and two sideband regions~(green). There are 42965 events in the signal region and 14728 in the combined sideband regions. }
\label{fig:show}
\end{figure}

The DT yields are determined by counting events in a signal and two sideband regions in the plane of $M_{D_s^+}$ versus $M_{D_s^-}$.  This provides a unified method that works well, considering the low statistics of some DT final states.
Figure~\ref{fig:show} shows the $M_{D_s^+}$ versus $M_{D_s^-}$ distribution of all DT candidates in data,
as well as the signal and the sideband  regions. 
The signal region requires that the average invariant mass satisfy $|\overline{m}-m_{D_s^+}|<15~\textrm{MeV}/c^2$ while the invariant mass difference $\Delta m = M_{D_s^+}-M_{D_s^-}$ satisfies $|\Delta m|<30~\textrm{MeV}/c^2$.
We define a sideband region with the same $\overline{m}$ requirement but with $80<|\Delta m|<140~\textrm{MeV}/c^2$.
The numbers of events for nineteen final states of seven data samples in the signal region and the sideband  regions are 42965 and 14728, respectively. 

The DT backgrounds have two main components: a uniform background 
and the crossfeed background. 
Based on the fact that the inclusive MC samples adequately describe the data sample, we estimate the uniform background by using inclusive MC samples to derive a scale factor $f$ for the uniform background as:
\begin{eqnarray}
\begin{aligned}
N_{\textrm{bkg}}=f \times N_{\textrm{sideband}},
\end{aligned}
\label{eq:f_DT}
\end{eqnarray}
where $N_{\textrm{bkg}}$ is the number of the uniform background events in the signal region
and $N_{\textrm{sideband}}$ is the number of the uniform background events in the sideband regions.
To describe the crossfeed background, 
we use a method analogous to the one used for the ST crossfeed to create an efficiency-correction matrix, $\textbf{C}^{\rm DT}$, 
and include this matrix in the BF fit.


\section{Systematic uncertainties}
\label{section:sys}

The systematic uncertainties in the BF measurements mainly come from
tracking and PID efficiencies, $\gamma$, $K_S^0$, $\pi^0$, and $\eta$ reconstruction, intermediate decays, amplitude model, MC statistics, background shape, crossfeed probabilities and the scale factor from the uniform background distribution of DT yields. 
The efficiency systematics accounts for the momentum distributions of the relevant particles.  We discuss each of them in detail below.

\begin{itemize}

\item $\pi^{\pm}, K^{\pm}$ tracking efficiency.
 The decays $e^+e^- \to K^+K^-K^+K^-$, $K^+K^-\pi^+\pi^-~(\pi^0)$, and $ \pi^+\pi^-\pi^+\pi^-~(\pi^0)$ are used to study the $K^+$ and $\pi^\pm$ tracking efficiencies. 
 The $\pi^+~(\pi^-)$ and $K^+~(K^-)$  data-MC tracking efficiency ratios, 
including the $\pi^{\pm}$ from $\eta_{\pi^+ \pi^- \pi^0}$, $\eta_{\pi^+ \pi^- \pi^0}'$ and $\rho(770)^0$, are given in table~\ref{tab:sys_1}. 
 \item $\pi^{\pm}, K^{\pm}$ PID efficiency.
The $\pi^{\pm} / K^{\pm}$ PID efficiency is studied with the same control samples as the tracking efficiency.
The $\pi^+~(\pi^-)$ and $K^+~(K^-)$ data-MC PID efficiency ratios,
including the $\pi^{\pm}$ from $\eta_{\pi^+ \pi^- \pi^0}$, $\eta_{\pi^+ \pi^- \pi^0}'$ and $\rho(770)^0$, are given in table~\ref{tab:sys_1}. 

 \item $K_S^0$ reconstruction.
The uncertainty for the $K_S^0$ reconstruction efficiency is studied using the control samples of ${\mathit J / \psi} \to K_S^0 K^+ \pi^-$ and $\phi K_S^0 K^+ \pi^-$ decays~\cite{sys_KS}. 
The data-MC $K_S^0$ reconstruction efficiency ratios are given in table~\ref{tab:sys_1}.

 \item $\pi^0$ and $\eta_{\gamma \gamma}$ reconstruction.
 The systematic uncertainty associated with the $\pi^0$ reconstruction efficiency is investigated by using a control sample of the process $e^+ e^- \to K^+ K^- \pi^+ \pi^- \pi^0$. 
 The systematic uncertainty for $\eta_{\gamma \gamma}$ reconstruction is assigned to be the same vs.~momentum as that of $\pi^0$ reconstruction.  
 The average ratio between data and MC efficiencies of $\pi^0$ and $\eta_{\gamma \gamma}$ reconstruction, weighted by the corresponding momentum spectra are given in table~\ref{tab:sys_1}. 

\begin{table}[!htbp]
\centering
\resizebox{\linewidth}{!}{
\begin{tabular}{l r@{$\pm$}l r@{$\pm$}l r@{$\pm$}l r@{$\pm$}l }
\hline
Final state  &  \multicolumn{2}{c}{Tracking} & \multicolumn{2}{c}{PID}  &  \multicolumn{2}{c}{$K_S^0$ }  & \multicolumn{2}{c}{$\pi^0 \to \gamma \gamma$~($\eta \to \gamma \gamma$)} \\
\hline
$K^{0}_{S} K^{+}$ & 1.001~ & ~0.002 & 1.000~ & ~0.002 & 1.020~ & ~0.005 & \multicolumn{2}{c}{-}\\
$K^{+} K^{-} \pi^{+}$ & 1.007~ & ~0.010 & 0.983~ & ~0.006 & \multicolumn{2}{c}{-} & \multicolumn{2}{c}{-}\\
$K^{0}_{S} K^{+} \pi^{0}$ & 1.002~ & ~0.002 & 0.996~ & ~0.002 & 1.028~ & ~0.006 & 1.014~ & ~0.010\\
$K^{0}_{S} K^{0}_{S} \pi^{+}$ & 0.995~ & ~0.004 & 0.995~ & ~0.002 & 1.056~ & ~0.013 & \multicolumn{2}{c}{-}\\
$K^{+} K^{-} \pi^{+} \pi^{0}$ & 1.010~ & ~0.015 & 0.976~ & ~0.006 & \multicolumn{2}{c}{-} & 1.017~ & ~0.010\\
$K^{0}_{S} K^{+} \pi^{+} \pi^{-}$ & 1.003~ & ~0.011 & 0.986~ & ~0.006 & 1.035~ & ~0.008 & \multicolumn{2}{c}{-}\\
$K^{0}_{S} K^{-} \pi^{+} \pi^{+}$ & 1.003~ & ~0.011 & 0.987~ & ~0.006 & 1.034~ & ~0.008 & \multicolumn{2}{c}{-}\\
$\pi^{+} \pi^{+} \pi^{-}$ & 1.005~ & ~0.006 & 0.991~ & ~0.006 & \multicolumn{2}{c}{-} & \multicolumn{2}{c}{-}\\
$\pi^{+} \eta$ & 1.002~ & ~0.002 & 0.994~ & ~0.002 & \multicolumn{2}{c}{-} & 0.987~ & ~0.016\\
$\pi^{+} \pi^{0} \eta$ & 0.999~ & ~0.002 & 0.996~ & ~0.002 & \multicolumn{2}{c}{-} & 0.996~ & ~0.020\\
$\pi^{+} \pi^{+} \pi^{-} \eta$ & 0.996~ & ~0.007 & 0.992~ & ~0.006 & \multicolumn{2}{c}{-} & 1.001~ & ~0.010\\
$\pi^{+} \eta_{3\pi}$ & 0.997~ & ~0.007 & 0.989~ & ~0.006 & \multicolumn{2}{c}{-} & 1.020~ & ~0.010\\
$\pi^{+} \eta' $ & 0.989~ & ~0.011 & 0.984~ & ~0.006 & \multicolumn{2}{c}{-} & 0.999~ & ~0.010\\
$\pi^{+} \pi^{0} \eta'$ & 0.980~ & ~0.013 & 0.981~ & ~0.006 & \multicolumn{2}{c}{-} & 1.025~ & ~0.020\\
$\pi^{+} \eta'_{\pi^{+}\pi^{-}\eta_{3\pi}}$ & 0.973~ & ~0.021 & 0.971~ & ~0.010 & \multicolumn{2}{c}{-} & 1.033~ & ~0.018\\
$\pi^{+} \eta'_{\rho \gamma}$ & 1.002~ & ~0.006 & 0.992~ & ~0.006 & \multicolumn{2}{c}{-} & \multicolumn{2}{c}{-}\\
$\pi^{+} \pi^{0} \eta'_{\rho \gamma}$ & 1.001~ & ~0.006 & 0.994~ & ~0.006 & \multicolumn{2}{c}{-} & 1.009~ & ~0.010\\
$K^{0}_{S} \pi^{+} \pi^{0}$ & 0.999~ & ~0.002 & 0.997~ & ~0.002 & 1.026~ & ~0.006 & 1.005~ & ~0.010\\
$K^{+} \pi^{+} \pi^{-}$ & 1.005~ & ~0.007 & 0.990~ & ~0.006 & \multicolumn{2}{c}{-} & \multicolumn{2}{c}{-}\\
\hline
\end{tabular}
}
\caption{The data-MC efficiency ratios. The MC efficiencies have been corrected to data by these ratios and the uncertainties of the ratios are assigned as the systematic uncertainties. A "-" indicates that the ratio is not applicable.}
\label{tab:sys_1}
\end{table}

 \item $\gamma$ reconstruction. The systematic uncertainty of $\gamma$ detection efficiency is 1$\%$ per photon, 
 obtained by studying the control sample of ${\mathit J / \psi} \to \rho(770)^0 \pi^0$~\cite{sys_gamma}.
 
  \item Intermediate resonance decays.
The uncertainties in the BFs of intermediate resonance decays are considered,
as listed in the PDG~\cite{PDG}: 
\begin{itemize}
    \item $\mathcal{B}~(K_S^0 \to \pi^+ \pi^-)=(69.20\pm0.05)\%$,
    \item $\mathcal{B}~(\pi^0 \to \gamma \gamma)=(98.823\pm0.034)\%$,
    \item $\mathcal{B}~(\eta \to \gamma \gamma)=(39.41\pm0.20)\%$,
    \item $\mathcal{B}~(\eta \to \pi^+ \pi^- \pi^0)=(22.68\pm0.23)\%$,
    \item $\mathcal{B}~(\eta' \to \pi^+ \pi^- \eta)=(42.5\pm0.5)\%$,
    \item $\mathcal{B}~(\eta' \to \rho(770)^0 \gamma)=(29.5\pm0.4)\%$.
\end{itemize}

 \item Amplitude model.
The uncertainties from the amplitude models are estimated by varying the amplitude model parameters based on their error matrix.
Here we assign the uncertainty according to the amplitude~\cite{mode_401,mode_402,mode_403,mode_404,mode_xxx,mode_406,mode_421,mode_441,mode_442,mode_461,mode_501,mode_502} as listed in table~\ref{tab:sys_2}.

\begin{table}[!htbp]
\centering
\begin{tabular}{l  c  c   c}
\hline 
\multirow{2}{*}{Final state}  & \multirow{2}{*}{$\gamma$ reconstruction}  & Intermediate   &  Amplitude \\
& ~& resonance decay & model \\
\hline
$  K^{0}_{S} K^{+}$                                           & -     & 0.07  & -       \\ 
$  K^{+} K^{-} \pi^{+}$                                       & -     & -     & 0.50   \\ 
$  K^{0}_{S} K^{+} \pi^{0}$                                   & -     & 0.08  & 0.80   \\ 
$  K^{0}_{S} K^{0}_{S} \pi^{+}$                               & -     & 0.14  & 0.50   \\ 
$  K^{+} K^{-} \pi^{+} \pi^{0}$                               & -     & 0.03  & 0.40   \\ 
$  K^{0}_{S} K^{+} \pi^{+} \pi^{-}$                           & -     & 0.07  & 0.40   \\ 
$  K^{0}_{S} K^{-} \pi^{+} \pi^{+}$                           & -     & 0.07  & 0.60   \\ 
$  \pi^{+} \pi^{+} \pi^{-}$                                   & -     & -     & 0.50   \\ 
$  \pi^{+} \eta_{\gamma \gamma}$                              & -     & 0.51  & -      \\ 
$  \pi^{+} \pi^{0} \eta_{\gamma \gamma}$                      & -     & 0.51  & 0.60   \\ 
$  \pi^{+} \pi^{+} \pi^{-} \eta_{\gamma \gamma}$              & -     & 0.51  & 0.40   \\ 
$  \pi^{+} \eta_{\pi^+ \pi^- \pi^0}$                          & -     & 1.01  & -      \\ 
$  \pi^{+} \eta_{\pi^+ \pi^- \eta_{\gamma \gamma}}' $         & -     & 1.28  & -      \\ 
$  \pi^{+} \pi^{0} \eta_{\pi^+ \pi^- \eta_{\gamma \gamma}}'$  & -     & 1.28  & 0.40   \\ 
$  \pi^{+} \eta'_{\pi^{+}\pi^{-}\eta_{\pi^+ \pi^- \pi^0}}$    & -     & 1.18  & -      \\ 
$  \pi^{+} \eta'_{\rho \gamma}$                               & 1.00  & 1.36  & -      \\ 
$  \pi^{+} \pi^{0} \eta'_{\rho \gamma}$                       & 1.00  & 1.36  & 0.50   \\ 
$  K^{0}_{S} \pi^{+} \pi^{0}$                                 & -     & 0.08  & 0.80   \\ 
$  K^{+} \pi^{+} \pi^{-}$                                     & -     & -     & 0.50   \\ 
\hline
\end{tabular}
\caption{Systematic uncertainties from $\gamma$ reconstruction, intermediate resonance decays and amplitude model for the BF measurement~(\%).  The "-" indicates that the uncertainty is not applicable.}
\label{tab:sys_2}
\end{table}

\item MC statistics.
The systematic uncertainties due to MC statistics arise from the statistical uncertainties of 266 ST and 2527 DT efficiencies. 
The total  uncertainty from MC statistics for each mode is given in table~\ref{tab:sys_3}. 

 \item Background shape.
 To estimate the uncertainty due to the background shape of the signal $D_s^+$ invariant mass distribution, 
the MC background shape is used to replace the second-order polynomial. 
The total  uncertainty of the background shape for each mode is given in table~\ref{tab:sys_3}. 

 \item Crossfeed probabilities.
The systematic uncertainty due to crossfeed probabilities is obtained by propagating the statistical uncertainties of the $\bf{C}^{\rm ST}$ and $\bf{C}^{\rm DT}$ matrices. 
The total  uncertainty from crossfeed probabilities for each mode is given in table~\ref{tab:sys_3}. 

 \item The scale factor from uniform background of DT yields.
The systematic uncertainty due to the scale factor from uniform background of DT yields is taken as the statistical uncertainty of $f$ from the inclusive MC sample.
The total  uncertainty of the scale factor from uniform background of DT yields for each mode is given in table~\ref{tab:sys_3}.

\begin{table}[!htbp]
\centering
\begin{tabular}{l  c  c  c  c }
\hline 
\multirow{2}{*}{Final state}  & MC & Background & Crossfeed & DT background \\
~ & statistics & shape & probabilities & factor \\
\hline
$ K^{0}_{S} K^{+}$ & 0.10   & 0.01   & 0.01   & 0.01  \\
$ K^{+} K^{-} \pi^{+}$ & 0.05   & 0.01   & 0.00   & 0.01  \\
$ K^{0}_{S} K^{+} \pi^{0}$ & 0.15   & 0.07   & 0.01   & 0.07  \\
$ K^{0}_{S} K^{0}_{S} \pi^{+}$ & 0.26   & 0.07   & 0.01   & 0.07  \\
$ K^{+} K^{-} \pi^{+} \pi^{0}$ & 0.08   & 0.04   & 0.00   & 0.04  \\
$ K^{0}_{S} K^{+} \pi^{+} \pi^{-}$ & 0.18   & 0.12   & 0.01   & 0.12  \\
$ K^{0}_{S} K^{-} \pi^{+} \pi^{+}$ & 0.14   & 0.05   & 0.00   & 0.05  \\
$ \pi^{+} \pi^{+} \pi^{-}$ & 0.09   & 0.06   & 0.01   & 0.06  \\
$ \pi^{+} \eta$ & 0.13   & 0.03   & 0.00   & 0.03  \\
$ \pi^{+} \pi^{0} \eta$ & 0.08   & 0.04   & 0.00   & 0.04  \\
$ \pi^{+} \pi^{+} \pi^{-} \eta$ & 0.16   & 0.13   & 0.01   & 0.13  \\
$ \pi^{+} \eta_{3\pi}$ & 0.25   & 0.05   & 0.01   & 0.05  \\
$ \pi^{+} \eta' $ & 0.18   & 0.02   & 0.00   & 0.02  \\
$ \pi^{+} \pi^{0} \eta'$ & 0.27   & 0.10   & 0.00   & 0.10  \\
$ \pi^{+} \eta'_{\pi^{+}\pi^{-}\eta_{3\pi}}$ & 0.50   & 0.05   & 0.00   & 0.05  \\
$ \pi^{+} \eta'_{\rho \gamma}$ & 0.12   & 0.05   & 0.00   & 0.05  \\
$ \pi^{+} \pi^{0} \eta'_{\rho \gamma}$ & 0.21   & 0.19   & 0.01   & 0.19  \\
$ K^{0}_{S} \pi^{+} \pi^{0}$ & 0.35   & 0.20   & 0.01   & 0.20  \\
$ K^{+} \pi^{+} \pi^{-}$ & 0.12   & 0.07   & 0.00   & 0.07  \\
\hline
\end{tabular}
\caption{Total systematic uncertainties from each of these sources: MC statistics, background shape, crossfeed probabilities and the scale factor from uniform background of DT yields for the BF measurement~(\%).}
\label{tab:sys_3}
\end{table}

\end{itemize}

For the data-MC efficiency ratios in table~\ref{tab:sys_1}, the MC efficiencies have been corrected by these ratios and the uncertainties of the ratios are assigned to be the systematic uncertainties. 


\section{Branching fraction measurement}
\label{section:fit}

We use a maximum-likelihood fit to obtain the BFs and the number of $D_s^+ D_s^-$ pairs by the observed ST and DT yields.


For the data sample $k$, 
we denote the number of $D_s^+ D_s^-$ pairs as $N^{D_s^+ D_s^-}_k$, 
the observed ST yield matrix as ${\bf{N} }^{\textrm{ST}}_k$ ($1 \times 38$ matrix) 
and the expected yield matrix ${\bf{Y}}^{\textrm{ST}}_k$~($38 \times 1$ matrix):
\begin{eqnarray}
\begin{aligned}
{\bf{Y}}^{\textrm{ST}}_k=N^{D_s^+ D_s^-}_k {\bf{\mathcal{B}}}^{\rm ST} {\bf{C}}^{{\rm ST}}_k,
\end{aligned}
\end{eqnarray}
where ${\bf{\mathcal{B}}}^{\rm ST}$~($1 \times 38$ matrix) is the BF matrix and 
${\bf{C}}^{{\rm ST}}_k$~($38 \times 38$ matrix) is the ST efficiency-correction matrix. 
The likelihood function of ST yields for data sample $k$~($\mathcal{L}^{\textrm{ST}}_k$) can be expressed as:
\begin{eqnarray}
\begin{aligned}
\mathcal{L}^{\textrm{ST}}_k \,=\, \frac{1}{\sqrt{2\pi \left|\det {\bf{V}}^{\rm ST}_k\right|}} \,
 {\rm exp} \left[ (
{\bf{Y}}^{\textrm{ST}}_k-{\bf{N}}^{\textrm{ST}}_k) 
({\bf{V}}^{\rm ST}_{k})^{-1}
 ({\bf{Y}}^{\textrm{ST}}_k-{\bf{N}}^{\textrm{ST}}_k)^T/2 
\right],
\end{aligned}
\end{eqnarray}
where 
$T$ is the transpose operation of a matrix, 
$\det$ is the determinant operation for a matrix, 
and 
${~\bf{V}}^{\rm ST}_k$~($38 \times 38$ matrix) is the ST statistical uncertainty matrix~\cite{sun2006simultaneous}:
\begin{eqnarray}
V_{k,ij}^{\rm ST}=
\left\{
\begin{aligned}
\sigma_{{\rm ST},i}\sigma_{{\rm ST},j}, ~& i=j \\
\sigma_{{\rm DT},ij}^2 ,~& i \neq j 
\end{aligned}
\right
.
\label{eq:VST}
\end{eqnarray}
The diagonal elements are the statistical uncertainty of ST yields~($\sigma_{{\rm ST},i}$) and the off-diagonal elements are evaluated as the observed DT yield~($\sigma_{{\rm DT},ij}$). 
Since any event can contain both ST and DT candidates, 
the ST yields are correlated among themselves as well as with DT yields.

For a DT final state, 
$D_s^+ \to i$ and $D_s^- \to \bar{j}$, 
in data sample $k$,
we denote the observed DT yield in sideband regions as $S^{\rm DT}_{i,\bar{j},k}$
and the observed DT yield in the signal region as $N^{\rm DT}_{i,\bar{j},k}$.
The expected yield at generator level is $\tilde{E}^{\rm DT}_{i,\bar{j},k}$:
\begin{eqnarray}
\begin{aligned}
\tilde{E}^{\rm DT}_{i,\bar{j},k}=N^{D_s^+ D_s^-}_k \mathcal{B}_{i} \mathcal{B}_{\bar{j}}.
\end{aligned}
\label{eq:Y_DT}
\end{eqnarray}
Based on Eq.~(\ref{eq:Y_DT}), 
we denote an expected yield matrix ${\tilde{\bf{E}}}^{\rm DT}_k$~($1 \times 361$ matrix) and an  efficiency-corrected expected yield matrix ${\bf{Y}}^{\rm DT}_k$~($1 \times 361$ matrix) as: 
\begin{eqnarray}
{\bf{Y}}^{\rm DT}_k={\tilde{\bf{E}}^{\rm DT}_k}  {\bf{C}}^{\rm DT}_k,
\end{eqnarray}
where ${\bf{C}}^{\rm DT}_k$~($361 \times 361$ matrix) is the DT efficiency-correction matrix constructed in Sec.~\ref{section:yields}.

The DT likelihood function $\mathcal{L}^{\rm DT}_{i,\bar{j},k}$ is given by the usual description of a Poisson signal in the presence of Poisson background~\cite{e2}:
\begin{eqnarray}
\begin{aligned}
&\mathcal{L}^{\rm DT}_{i,\bar{j},k} (N^{\rm DT}_{i,\bar{j},k},S^{\rm DT}_{i,\bar{j},k};Y^{\rm DT}_{i,\bar{j},k},\tilde{S}^{\rm DT}_{i,\bar{j},k})=\\
&\frac{1}{(N^{\rm DT}_{i,\bar{j},k})!} \frac{ (Y^{\rm DT}_{i,\bar{j},k}+f_{i,\bar{j},k} \tilde{S}^{\rm DT}_{i,\bar{j},k})^{N^{\rm DT}_{i,\bar{j},k}}}{{\rm exp} (Y^{\rm DT}_{i,\bar{j},k}+f_{i,\bar{j},k} \tilde{S}^{\rm DT}_{i,\bar{j},k})}
 \frac{1}{ (f_{i,\bar{j},k}  S^{\rm DT}_{i,\bar{j},k})!} \frac{ (f_{i,\bar{j},k} \tilde{S}^{\rm DT}_{i,\bar{j},k})^{(f_{i,\bar{j},k} S^{\rm DT}_{i,\bar{j},k})}}
{{\rm exp} (f_{i,\bar{j},k} \tilde{S}^{\rm DT}_{i,\bar{j},k})}
,
\end{aligned}
\label{eq:likelihood_P}
\end{eqnarray}
where $f_{i,\bar{j},k}$ is the scale factor from uniform background of DT yields in Eq.~(\ref{eq:f_DT}) 
and ${\tilde{S}}^{\rm DT}_{i,\bar{j},k}$ is the expected yield in sideband regions and eliminated by solving ${\partial (\mathcal{L}^{\rm DT}_{i,\bar{j},k})}/{\partial {\tilde{S}}^{\rm DT}_{i,\bar{j},k}}=0$ analytically.

Based on seven data samples and $19 \times 19$ DT final states,
the final likelihood function~($\mathcal{L}$) can be expressed as:
\begin{eqnarray}
\begin{aligned}
\mathcal{L}=\prod_{k=1}^{7}~\mathcal{L}^{\rm ST}_{k} \prod_{i=1}^{19} \prod_{\bar{j}=1}^{19} \mathcal{L}^{\rm DT}_{i,\bar{j},k}.
\end{aligned}
\end{eqnarray}
The results of the BF fit and comparisons to the PDG~\cite{PDG} are listed in table~\ref{tab:result}.
The fitted numbers of produced $D_s^+ D_s^-$ pairs for
 seven data samples are list in table~\ref{tab:ndsds}.

\begin{table}[!htbp]
 \centering
 \begin{tabular}{l c r@{$\pm$}l}
 \hline 
 Mode                                           & $\mathcal{B}~(\%)$         & \multicolumn{2}{c}{PDG $\mathcal{B}~(\%)$}  \\
\hline 
$D_{s}^{+} \to K^{0}_{S} K^{+}$                 & $ 1.502 \pm 0.012 \pm 0.009  $ & $  1.453 ~$ & $~ 0.035 $          \\
$D_{s}^{+} \to K^{+} K^{-} \pi^{+}$             & $ 5.49 \pm 0.04 \pm 0.07  $ & $  5.37 ~$ & $~ 0.10 $         \\
$D_{s}^{+} \to K^{0}_{S} K^{+} \pi^{0}$         & $ 1.47 \pm 0.02 \pm 0.02  $ & $  1.47 ~$ & $~ 0.07 $          \\
$D_{s}^{+} \to K^{0}_{S} K^{0}_{S} \pi^{+}$     & $ 0.73 \pm 0.01 \pm 0.01  $ & $  0.71 ~$ & $~ 0.04 $         \\
$D_{s}^{+} \to K^{+} K^{-} \pi^{+} \pi^{0}$     & $ 5.50 \pm 0.05 \pm 0.11  $ & $  5.50 ~$ & $~ 0.24 $          \\
$D_{s}^{+} \to K^{0}_{S} K^{+} \pi^{+} \pi^{-}$ & $ 0.93 \pm 0.02 \pm 0.01  $ & $  0.95 ~$ & $~ 0.08 $         \\
$D_{s}^{+} \to K^{0}_{S} K^{-} \pi^{+} \pi^{+}$ & $ 1.56 \pm 0.02 \pm 0.02  $ & $  1.53 ~$ & $~ 0.08 $         \\
$D_{s}^{+} \to \pi^{+} \pi^{+} \pi^{-}$         & $ 1.09 \pm 0.01 \pm 0.01  $ & $  1.08 ~$ & $~ 0.04 $         \\
$D_{s}^{+} \to \pi^{+} \eta$                    & $ 1.69 \pm 0.02 \pm 0.02  $ & $  1.67 ~$ & $~ 0.09 $         \\
$D_{s}^{+} \to \pi^{+} \pi^{0} \eta$            & $ 9.10 \pm 0.09 \pm 0.15  $ & $  9.5  ~$ & $~ 0.5  $                    \\
$D_{s}^{+} \to \pi^{+} \pi^{+} \pi^{-} \eta$    & $ 3.08 \pm 0.06 \pm 0.05  $ & $  3.12 ~$ & $~ 0.16 $             \\
$D_{s}^{+} \to \pi^{+} \eta' $                  & $ 3.95 \pm 0.04 \pm 0.07  $ & $  3.94 ~$ & $~ 0.25 $         \\
$D_{s}^{+} \to \pi^{+} \pi^{0} \eta'$           & $ 6.17 \pm 0.12 \pm 0.14  $ & $  6.08  ~$ & $~ 0.29  $              \\
$D_{s}^{+} \to K^{0}_{S} \pi^{+} \pi^{0}$       & $ 0.51 \pm 0.02 \pm 0.01  $ & $  0.54 ~$ & $~ 0.03 $               \\
$D_{s}^{+} \to K^{+} \pi^{+} \pi^{-}$           & $ 0.620 \pm 0.009 \pm 0.006  $ & $  0.620 ~$ & $~0.019 $        \\
 \hline
 \end{tabular}
      \caption{Results of the BF fit and comparison to the PDG values. 
      For results in this paper, the first uncertainties are statistical and the second ones systematic. 
      For the PDG total uncertainties are shown. }
  \label{tab:result}
\end{table}

\begin{table}[!htbp]
 \centering
 \begin{tabular}{c r@{$\pm$}c@{$\pm$}l}
 \hline 
 $\sqrt{s}~({\rm GeV})$ & \multicolumn{3}{c}{$N^{D_s^+ D_s^-}~(\times 10^5)$}\\
\hline
4.128 and 4.157 &             6.29 & 0.06 & 0.01 \\    
4.178           &            31.79 & 0.24 & 0.06 \\
4.189           &             5.51 & 0.05 & 0.01 \\
4.199           &             4.92 & 0.05 & 0.01 \\
4.209           &             5.07 & 0.05 & 0.01 \\
4.219           &             4.32 & 0.04 & 0.01 \\
4.226           &             6.82 & 0.07 & 0.02 \\
      \hline
 \end{tabular}
 \caption{The numbers of produced $D_s^+ D_s^-$ pairs for the seven data samples. The first and second uncertainties are statistical and systematic uncertainties, respectively. }
  \label{tab:ndsds}
\end{table}


We check the internal consistency of the BF fitting procedure using the inclusive MC sample, 
which corresponds to an integrated luminosity of 40 times the recorded data set, 
and find that central values and pull distributions for the fitted parameters are reasonable. 


Systematic uncertainties are propagated to the final results by adjusting fit inputs, including efficiencies, ST yields, DT yields in the signal region and the sideband regions and scale factors from uniform background of DT yields.
The appropriate correlations are included using the detailed results
of Sec.~\ref{section:sys}.  
The uncertainties from tracking and PID efficiencies, intermediate decays, amplitude model and the $K_S^0$, $\pi^0$, $\eta$, and $\gamma$ reconstruction are correlated while the uncertainties from MC statistics, ST background shape, crossfeed probabilities and the background of DT yields are uncorrelated. 

For each correlated systematic uncertainty,
we change all corresponding input values of all ST final states and DT final states with the expected systematic in that mode and obtain the new fit values.
For each uncorrelated systematic uncertainty from every final state,
we only change the corresponding input value of that final state with
the expected systematics.
For each parameter, the changes in the fitted value are taken as the contribution to the systematic uncertainty, and the sum of all contributions in quadrature gives the total systematic uncertainty.


\section{\emph{CP} asymmetries measurement}
\label{sec:07}

We derive the {\emph{CP}} asymmetry  for each mode by:
\begin{eqnarray}
\begin{aligned}
\mathcal{A}_{\emph{CP},i}=\frac
{\sum_{k=1}^7  (N^{\textrm{ST}}_{i,k}/\epsilon^{\textrm{ST}}_{i,k}-N^{\textrm{ST}}_{\bar{i},k}/\epsilon^{\textrm{ST}}_{\bar{i},k})}
{\sum_{k=1}^7  (N^{\textrm{ST}}_{i,k}/\epsilon^{\textrm{ST}}_{i,k}+N^{\textrm{ST}}_{\bar{i},k}/\epsilon^{\textrm{ST}}_{\bar{i},k})},
\end{aligned}
\end{eqnarray}
where $N^{\textrm{ST}}_{i,k}$ and $\epsilon^{\textrm{ST}}_{i,k}$~($N^{\textrm{ST}}_{\bar{i},k}$ and $\epsilon^{\textrm{ST}}_{\bar{i},k}$) are the yields and efficiencies for the ST mode $i$~($\bar{i}$) from date sample $k$.
Almost all systematic uncertainties cancel in $\mathcal{A}_{\emph{CP}}$ calculations,
except the ST efficiency statistical uncertainty, the ST yield fit uncertainties, and the uncertainties from tracking efficiencies.
The results of the $\mathcal{A}_{\emph{CP}}$ calculation and comparison to the PDG~\cite{PDG} are listed in table~\ref{tab:result_acp}.
\begin{table}[!htbp]
 \centering
 \begin{tabular}{l r@{$\,\pm\,$}c@{$\,\pm\,$}l r@{$\,\pm\,$}l}
 \hline 
 Mode & \multicolumn{3}{c}{$\mathcal{A}_{\emph{CP}}~(\%)$} & \multicolumn{2}{c}{PDG $\mathcal{A}_{\emph{CP}}~(\%)$}  \\
 \hline
$D_{s}^{\pm} \to K^{0}_{S} K^{\pm}$                              &     0.29 & 0.50 & 0.21 &     0.09  &  0.26          \\
$D_{s}^{\pm} \to K^{+} K^{-} \pi^{\pm}$                          &     0.48 & 0.26 & 0.24 &  $-$0.5   &  0.9           \\
$D_{s}^{\pm} \to K^{0}_{S} K^{\pm} \pi^{0}$                      &  $-$0.85 & 1.97 & 0.46 &  $-$2     &  6             \\
$D_{s}^{\pm} \to K^{0}_{S} K^{0}_{S} \pi^{\pm}$                  &     1.14 & 1.58 & 0.44 &     3     &  5             \\
$D_{s}^{\pm} \to K^{+} K^{-} \pi^{\pm} \pi^{0}$                  &  $-$0.66 & 0.91 & 0.33 &     0.0   &  3.0           \\
$D_{s}^{\pm} \to K^{0}_{S} K^{\pm} \pi^{+} \pi^{-}$              &     2.00 & 2.37 & 0.70 &  $-$6     &  5             \\
$D_{s}^{\pm} \to K^{0}_{S} K^{\mp} \pi^{\pm} \pi^{\pm}$          &  $-$0.24 & 1.05 & 1.07 &     4.1   &  2.8           \\
$D_{s}^{\pm} \to \pi^{\pm} \pi^{+} \pi^{-}$                      &  $-$0.88 & 1.17 & 0.38 &  $-$0.7   &  3.1           \\
$D_{s}^{\pm} \to \pi^{\pm} \eta$                                 &  $-$0.44 & 0.89 & 0.19 &     0.3   &  0.4           \\
$D_{s}^{\pm} \to \pi^{\pm} \pi^{0} \eta$                         &     1.05 & 1.45 & 0.62 &  $-$1     &  4             \\
$D_{s}^{\pm} \to \pi^{\pm} \pi^{+} \pi^{-} \eta$                 &     2.42 & 2.85 & 0.78 &  \multicolumn{2}{c}{-}     \\
$D_{s}^{\pm} \to \pi^{\pm} \eta' $                               &  $-$0.59 & 0.76 & 0.20 &  $-$0.9   & 0.5           \\
$D_{s}^{\pm} \to \pi^{\pm} \pi^{0} \eta'$                        &  $-$1.60 & 2.57 & 0.64 &     0     &  8             \\
$D_{s}^{\pm} \to K^{0}_{S} \pi^{\pm} \pi^{0}$                    &  $-$2.17 & 4.65 & 1.10 &     3     &  6            \\
$D_{s}^{\pm} \to K^{\pm} \pi^{+} \pi^{-}$                        &     1.81 & 2.01 & 0.45 &     4     & 5             \\
 \hline
 \end{tabular}\\
 \caption{Results of the $\mathcal{A}_{\emph{CP}}$ calculation for this paper and comparison to the PDG. For results in this paper, the first uncertainties are statistical and the second ones systematic.  For the PDG total uncertainties are shown.}
 \label{tab:result_acp}
\end{table}


\section{Summary}

We have measured the absolute BFs for fifteen hadronic $D_s^+$ decays, 
reconstructed in nineteen final states,  
using a sample of $e^+ e^-$ collision data  corresponding to an integrated luminosity of 7.33 $\rm{fb}^{-1}$ collected with the BESIII detector at center-of-mass energies between 4.128 and 4.226~GeV.
The BFs obtained and shown in table~\ref{tab:result} 
are in agreement with the world-average values~\cite{PDG}, 
but typically with much improved precision. 
The BFs of selected important reference modes are: 
\begin{equation*}
    \mathcal{B}(D_s^+ \to K^+ K^- \pi^+)=(5.49 \pm 0.04 \pm 0.07)\%, 
\end{equation*}
\begin{equation*}
    \mathcal{B}(D_s^+ \to K_S^0 K^+)=(1.50 \pm 0.01 \pm 0.01)\%,
\end{equation*}
\begin{equation*}
    \mathcal{B}(D_s^+ \to K^+ K^- \pi^+ \pi^0)=(5.50 \pm 0.05 \pm 0.11)\%,
\end{equation*}
where the first uncertainties are statistical and the second are systematic.
Additionally, the \emph{CP}-violating asymmetries of the fifteen hadronic $D_s^{\pm}$ decays are measured.  No significant asymmetries are observed.


\textbf{Acknowledgement}

The BESIII Collaboration thanks the staff of BEPCII and the IHEP computing center for their strong support. This work is supported in part by National Key R\&D Program of China under Contracts Nos. 2020YFA0406400, 2020YFA0406300; National Natural Science Foundation of China (NSFC) under Contracts Nos. 11635010, 11735014, 11835012, 11935015, 11935016, 11935018, 11961141012, 12025502, 12035009, 12035013, 12061131003, 12192260, 12192261, 12192262, 12192263, 12192264, 12192265, 12221005, 12225509, 12235017; the Chinese Academy of Sciences (CAS) Large-Scale Scientific Facility Program; the CAS Center for Excellence in Particle Physics (CCEPP); Joint Large-Scale Scientific Facility Funds of the NSFC and CAS under Contract No. U1832207; CAS Key Research Program of Frontier Sciences under Contracts Nos. QYZDJ-SSW-SLH003, QYZDJ-SSW-SLH040; 100 Talents Program of CAS; The Institute of Nuclear and Particle Physics (INPAC) and Shanghai Key Laboratory for Particle Physics and Cosmology; European Union's Horizon 2020 research and innovation programme under Marie Sklodowska-Curie grant agreement under Contract No. 894790; German Research Foundation DFG under Contracts Nos. 455635585, Collaborative Research Center CRC 1044, FOR5327, GRK 2149; Istituto Nazionale di Fisica Nucleare, Italy; Ministry of Development of Turkey under Contract No. DPT2006K-120470; National Research Foundation of Korea under Contract No. NRF-2022R1A2C1092335; National Science and Technology fund of Mongolia; National Science Research and Innovation Fund (NSRF) via the Program Management Unit for Human Resources \& Institutional Development, Research and Innovation of Thailand under Contract No. B16F640076; Polish National Science Centre under Contract No. 2019/35/O/ST2/02907; The Swedish Research Council; U. S. Department of Energy under Contract No. DE-FG02-05ER41374.

\bibliographystyle{JHEP}
\bibliography{references}

\clearpage
\appendix
M.~Ablikim$^{1}$, M.~N.~Achasov$^{4,c}$, P.~Adlarson$^{75}$, O.~Afedulidis$^{3}$, X.~C.~Ai$^{80}$, R.~Aliberti$^{35}$, A.~Amoroso$^{74A,74C}$, Q.~An$^{71,58,a}$, Y.~Bai$^{57}$, O.~Bakina$^{36}$
% ...
\\
\vspace{0.2cm}
(BESIII Collaboration)\\
\vspace{0.2cm} {\it
$^{1}$ Institute of High Energy Physics, Beijing 100049, People's Republic of China\\
% ...
$^{3}$ Bochum  Ruhr-University, D-44780 Bochum, Germany\\
$^{4}$ Budker Institute of Nuclear Physics SB RAS (BINP), Novosibirsk 630090, Russia\\
% ...
$^{35}$ Johannes Gutenberg University of Mainz, Johann-Joachim-Becher-Weg 45, D-55099 Mainz, Germany\\
$^{36}$ Joint Institute for Nuclear Research, 141980 Dubna, Moscow region, Russia\\
% ...
$^{57}$ Southeast University, Nanjing 211100, People's Republic of China\\
$^{58}$ State Key Laboratory of Particle Detection and Electronics, Beijing 100049, Hefei 230026, People's Republic of China\\
% ...
$^{71}$ University of Science and Technology of China, Hefei 230026, People's Republic of China\\
% ...
$^{74}$ University of Turin and INFN, (A)University of Turin, I-10125, Turin, Italy; (B)University of Eastern Piedmont, I-15121, Alessandria, Italy; (C)INFN, I-10125, Turin, Italy\\
$^{75}$ Uppsala University, Box 516, SE-75120 Uppsala, Sweden\\
% ...
$^{80}$ Zhengzhou University, Zhengzhou 450001, People's Republic of China\\

\vspace{0.2cm}
$^{a}$ Deceased\\
% ...
$^{c}$ Also at the Novosibirsk State University, Novosibirsk, 630090, Russia\\
% ...

}


\end{document}